\documentclass{mem}
\usepackage{natbib}\usepackage{txfonts}\usepackage{balance}
\usepackage{graphicx}
\usepackage[a4paper]{hyperref}
\idline{75}{282}
\newcommand{\lsim}{\ \raise -2.truept\hbox{\rlap{\hbox{$\sim$}}\raise
5.truept\hbox{$<$}\ }}
\newcommand{\gsim}{\ \raise -2.truept\hbox{\rlap{\hbox{$\sim$}}\raise
5.truept\hbox{$>$}\ }} \newcommand{\grv}{\`}

\begin{document}

\title{Optical SBFs of shell galaxies} 
\subtitle{} 

\author{ Biscardi I. \inst{1,2}, Raimondo
G.\inst{1}, Cantiello, M.\inst{1,3}, Brocato, E.\inst{1}}

\institute{INAF-Osservatorio Astronomico di Teramo, Via M.  Maggini
s.n.c., I-64100 Teramo, Italy.  \and
Dipartimento di Fisica - Universit\grv a di Roma Tor Vergata, via
della Ricerca Scientifica 1, 00133 Rome, Italy. 
 \newline \email{biscardi@oa-teramo.inaf.it}}
 
\offprints{I. Biscardi}
\authorrunning{Biscardi }

\titlerunning{Optical SBF in shell galaxies}

\abstract {We measure $F814W$ Surface Brightness Fluctuations (SBFs)
for a sample of distant shell galaxies observed with the Advanced
Camera for Survey (ACS) on board of HST.  To evaluate the distance at
galaxies, theoretical SBF magnitudes for the ACS@HST filters are
computed for single burst stellar populations covering a wide range of
ages (t=1.5-14 Gyr) and metallicities (Z=0.008-0.04).  Using these
stellar population models we provide the first $\bar{M}_{F814W}$
versus $(F475W-F814W)_0$ calibration.  The results suggest that
\emph{optical} SBFs can be measured at $d \geq 100 Mpc$ using high
resolution spatial optical data.

}

\maketitle{}

\section{Introduction}
The SBF method is a powerful technique to derive distance to galaxies
as far as 130 Mpc with uncertainties lower then 10\%.  The SBFs can be
evaluated for elliptical, spiral galaxies with prominent bulge and
dwarf galaxies.  An interesting case is represented by shell
elliptical galaxies.  Shell structures are considered robust
indicator of past interactions events and the stellar population in
the shell depends on the galaxy with which the merger has taken place
(e.g. \citet{Malin&Carter83}) and on the time spent since the shell
structure has formed. Then, it is reasonable to expect that the presence of
shells might influence the SBF signal of the galaxy. In this respect,
the high quality of ACS images is crucial to allow the measurement of
SBF of the galaxy, even at high distances.

\section{Data Analysis}
To derive the SBF measurements from ACS images, we follow the  procedure
adopted in \citet{Cantiello+05,Cantiello+07a} and \citet{Biscardi+08}, that we can
summarize as follow:
\begin{enumerate}
\item Sky and galaxy-model subtraction.
\item Removal of large residual background + shell features (residual
frames determination).
\item Photometry of external sources (globular cluster + background galaxies).
\item Mask of the most prominent unsubracted shell, if necessary.
\item Evaluation of the residual frame power spectrum, and its fitting
using a PSF template power spectrum. 
\end{enumerate}

\section{Stellar Population models and Calibrations}
\begin{figure}[!t]
\begin{center}
\resizebox{5cm}{!}{\includegraphics[clip=true]{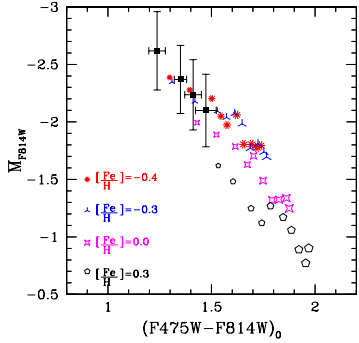}}
 \caption {\footnotesize SSP models of different metallicities (as
labeled) and ages (t=1.5-14 Gyr). Symbols with increasing size mark
models of older age. Black squares are observed galaxies }
 \label{fig:fig1}
\end{center}
\end{figure}

\begin{figure}[t!]
\begin{center}
\resizebox{5cm}{!}{\includegraphics[clip=true]{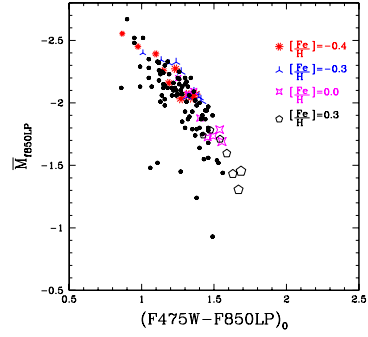}}
 \caption {\footnotesize The new SBF models in the ACS photometric
system, shifted to a Virgo distance modulus of 31.1 (symbols are as in
Fig. \ref{fig:fig1}) with observational data (full dots) are from
ACS Virgo Survey \citet{Mei+07}.}
 \label{fig2}
\end{center}
\end{figure}

We provide new SBF predictions together with new calibrations of
absolute SBF magnitudes for the ACS HST filters. These models, showed in
Figure \ref{fig:fig1},  are the most updated version of the code
SPoT \citep[Stellar Population
Tools,][]{Raimondo+05b}\footnote{http://www.oa-teramo.inaf.it/SPoT} and they 
are based on single-burst stellar population (SSP) of age ranging from
t=1.5 Gyr up to t=14 Gyr and metallicity from Z=0.008 to Z=0.04.
Using these SSP models we derive the following theoretical calibrations:
\scriptsize\begin{equation}
\label{eq:calibACS1}
\overline{M}_{F814W}= (-0.94 \pm 0.20)+(2.2 \pm0.2) \times [(F475W- F814W)_{0} - 2.0]
\end{equation}
\begin{equation}
\label{eq:calibMEI}
\overline{M}_{F850LP}= (-2.1 \pm 0.2)+
(1.4 \pm0.2) \times [(F475W- F850LP)_{0} - 1.3]
\end{equation}
\begin{equation}
\footnotesize
\label{eq:myjensen} 
\overline{M}_{I}= (-1.6 \pm 0.1)+ (4.5 \pm0.2)\times [(V-I)_{0}-1.15]
\end{equation}
\normalsize
which provide a very good agreement when compared with observational
data (Figure \ref{fig2}).
\section{ Distance and H$_{0}$ determination}
Coupling theoretical calibration (Eq. \ref{eq:calibACS1}) and $F814W$
 SBF measurements, the distance moduli are derived for the first time
 for the galaxies in Table \ref{tab1}. The present measurements show
 that distances of galaxies beyond 100 Mpc can be derived in optical
 filters, with SBF method.
\begin{table}[t!]
\caption{Distance Moduli estimated in this work.}
\begin{center}
\begin{tabular}{cc}
\hline 
Galaxy & $DM$ \\ 
\hline 
PGC\,6510 & 33.7 $\pm$ 0.25 \\
 PGC\,10922 & 34.2 $\pm$ 0.25 \\ PGC\,42871 & 34.7 $\pm$ 0.25 \\
 PGC\,6240 & 35.2 $\pm$ 0.25 \\ 
\hline
\label{tab1}
\end{tabular}
\end{center}
\end{table}

 An estimation of $H_{0} = 76 \pm 6$
 $\pm 5 \ km/s/Mpc$ is also obtained.

\bibliographystyle{aa}

\end{document}